\documentclass[sensors,article,submit,pdftex,moreauthors]{Definitions/mdpi} 
\firstpage{1} 
\pubvolume{1}
\issuenum{1}
\articlenumber{0}
\pubyear{2022}
\copyrightyear{2022}
\datereceived{} 
\dateaccepted{} 
\datepublished{} 
\hreflink{https://doi.org/} % If needed use \linebreak
\usepackage{amsfonts,amssymb}
\graphicspath{{./figures/}}
\usepackage{subcaption}
\usepackage{siunitx}
\ifdefined\qtyproduct
\else
    \ifdefined\NewCommandCopy
        \NewCommandCopy\qtyproduct\SI
    \else
        \NewDocumentCommand\qtyproduct{O{}mm}{\SI[#1]{#2}{#3}}
    \fi
\fi
\newcommand{\var}[1]{{\operatorname{\mathit{#1}}}}

\usepackage{algpseudocode}
\usepackage[vlined, linesnumbered]{algorithm2e}
\makeatletter
\renewcommand{\@algocf@capt@plain}{above}% formerly {bottom}
\newcommand{\nosemic}{\renewcommand{\@endalgocfline}{\relax}}% Drop semi-colon ;
\newcommand{\dosemic}{\renewcommand{\@endalgocfline}{\algocf@endline}}% Reinstate semi-colon ;
\makeatother

\preto{\abstractkeywords}{\nolinenumbers}

\Title{A Bayesian Method for Material Identification of Composite Plates via Dispersion Curves}

\TitleCitation{A Bayesian Method for Material Identification of Composite Plates via Dispersion Curves}

\Author{Marcus Haywood-Alexander $^{1}$*, Nikolaos Dervilis $^{1}$, Keith Worden $^{1}$, Robin S.\ Mills $^{2}$, Purim Ladpli $^{3}$ and Timothy J.\ Rogers $^{1}$}

\AuthorNames{Marcus Haywood-Alexander, Nikolaos Dervilis, Keith Worden, Robin S.\ Mills, Purim Ladpli and Timothy J.\ Rogers}

\AuthorCitation{Haywood-Alexander, M.; Dervilis, N.; Worden, K.; Mills, R.S.; Ladpli, P.; and Rogers, T.J.}
\address{%
$^{1}$ \quad Dynamics Research Group, Department of Mechanical Engineering, Mappin Street, The University of Sheffield, Sheffield, S1 3JD, UK\\
$^{2}$ \quad Dynamics Research Group, Laboratory for Verification and Validation (LVV), Europa Avenue, Sheffield, S9 1ZA, UK\\
$^{3}$ \quad Siemens Gamesa Renewable Energy, Assensjev 11, 9220 Aalborg, Denmark}

\corres{Correspondence: m.haywood@sheffield.ac.uk}

\abstract{Ultrasonic guided waves offer a convenient and practical approach to structural health monitoring and non-destructive evaluation. %
A key property of guided waves is the fully-defined relationship between central frequency and propagation characteristics (phase velocity, group velocity and wavenumber) -- which is described using dispersion curves. %
For many guided wave-based strategies, accurate dispersion curve information is invaluable, such as group velocity for localisation. %
From experimental observations of dispersion curves, a system identification procedure can be used to determine the governing material properties. %
As well as returning an estimated value, it is useful to determine the distribution of these properties based on measured data. %
A method of simulating samples from these distributions is to use the iterative Markov-Chain Monte Carlo (MCMC) procedure, which allows for freedom in the shape of the posterior. %
In this work, a scanning-laser doppler vibrometer is used to record the propagation of Lamb waves in a unidirectional-glass-fibre composite plate, and dispersion curve data for various propagation angles are extracted. %
Using these measured dispersion curve data, the MCMC sampling procedure is performed to provide a Bayesian approach to determining the dispersion curve information for an arbitrary plate. %
The distribution of the material properties at each angle is discussed, including the inferred confidence in the predicted parameters.}


\begin{document}

%%%%%%%%%%%%%%%%%%%%%%%%%%%%%%%%%%%%%%%%%%

\section{Introduction}

This paper focusses on determining dispersive characteristics of Lamb waves in arbitrary plates, with an emphasis on their use in non-destructive evaluation (NDE) and structural health monitoring (SHM). %
The use of ultrasonic guided waves (UGWs) for SHM strategies \cite{Rose2004} can offer a number of distinct advantages, such as range and sizing potential, greater sensitivity and cost effectiveness. %
There are, commonly, three types of high-frequency stress waves which fall under the category of guided waves: Rayleigh waves, Lamb waves and shear horizontal waves. %
The former of these types propagate on a surface, whereas the latter propagate in `thin' plates. %
Full descriptions and derivations of Rayleigh and Lamb waves are in \cite{Viktorov1967,Worden2001,Rose2014}, although a short introduction to some key concepts will be given here. %
A particular distinction of Lamb waves is their separation into \emph{symmetric} modes, which have the upper and lower plate surfaces oscillating in opposite directions at equal propagation distance, and \emph{antisymmetric} modes, with the oscillations in the same direction. %
For these two wave modes, the oscillation direction is perpendicular to the wave-guide surface, the \emph{shear horizontal} modes oscillate in the same direction as the propagation, and the solutions to these are often obtained along with solutions for Lamb waves. %
Higher-order modes will also be present with an increased \emph{frequency-thickness} product. %
When a Lamb wave is actuated in a plate, multiple wave modes will propagate from the source, of varying frequencies and shapes. %
As the propagation velocity of these waves depends on the central frequency of the wave and its shape, a wave-packet of mixed wavelengths with spread out in space; i.e., it will \emph{disperse}. %

This relationship is more completely described by defining a map between the frequency and wavenumber, which can be plotted as \emph{dispersion curves}. %
Use of dispersion curve information is essential in guided wave-based NDE and SHM strategies \cite{alleyne1992optimization,cawley1996use,guo1994lamb}, one example being to use known group velocities for damage localisation \cite{Haywood2021_location,Kundu2014}. %
In practice, the governing elastodynamic equations are numerically solved to determine these curves. %
For isotropic materials, this can be done using a simple iterative procedure to find the phase velocity at a given frequency \cite{Rose2014}. %

However, modelling guided-wave phenomena in complex materials is much more difficult than for isotropic materials, thanks to their anisotropy resulting in more complicated phenomena, as well as the need for a larger quantity of material properties. %
For more complex materials, there is no standard method of solving dispersion curves, although many are available which have distinct advantages for different uses. %
An approach by Solie and Auld \cite{Solie1973}, attempts to derive the equations using the partial-wave technique. %
This method assumes that the Lamb wave can be formulated as the superposition of three upward and three downward waves, each of which is referred to as a `partial wave'. %
Traditionally, matrix formulations are also used to retrieve wave propagation characteristics for a given frequency \cite{Kundu2019}. %

Further examples of finite element methods to model dispersion curves are shown by Shorter \cite{shorter2004wave}, or Manconi and Sorokin \cite{manconi2013effect}, which uses complex velocities to model viscoelastic behaviour of the material. %
An improvement in computational efficiency was made on these by using a semi-analytical-finite-element (SAFE) method \cite{Fan2010Thesis,yang2020guided}. %
Another computationally efficient method of calculating dispersion-curve solutions is the spectral element method, which uses Chebyshev polynomials to form an eigenvalue problem \cite{xiao2016guided}, which requires manipulation of material properties to form the constitutive matrices \cite{moll2010multi}. %

The final method of dispersion curve solutions to be mentioned here is the Legendre polynomial expansion approach first shown by Lefebvre \cite{Lefebvre2001}, which utilises the orthonormal basis set to form an eigenvalue problem. %
Over time, methods of solutions to dispersion curves have increased in computational efficiency, without significant loss in accuracy. %
This improvement, along with the increase in available computation power, is opening the door for identification procedures. %

From the governing equations, the dispersion curves are defined by a list of material properties, the number of which can become extensive for anisotropic and/or inhomogeneous materials. %
It follows then, that information on the dispersion curves may allow for inference of these material properties. %
Eremin \cite{eremin2015evaluation} showed how orthotropic material properties could be found using a genetic algorithm, where they minimised an objective function based on full experimental image data. %
An alternative method which included the complex wavenumber was shown by Roozen \cite{roozen2017estimation}, using Hankel's function to reconstruct the full wavefield in an isotropic plate. %
Work has also been shown by Webersen \cite{webersen2018guided}, which uses the SAFE method to reconstruct the dispersion curves, and the parameters are estimated by minimisation of an objective function. %
An obstacle of using dispersion curves for material identification is the difficult, or impossibility, of an analytical or numerical inversion of the solution methods \cite{cui2019identification}. %
One method of overcoming this is to use machine learning methods such as neural networks \cite{rautela2020ultrasonic,gopalakrishnan2020deep}, or genetic algorithms \cite{kudela2021elastic}. %

As well as an estimation of the most likely values, it is also useful to determine the posterior distribution of these parameters. %
Some advantages of estimating these distributions include accounting for environmental conditions and for uncertainty propagation. %
A laudable example of this has been shown, using a genetic algorithm and extending the list of parameters to include a noise term \cite{kudela2020elastic}; this generated feasible elastic constants and a distribution based on an assumed Gaussian posterior. %
However, this assumption of the posterior shape is a shortcoming of the approach, as well as the absence of any possible inference on the cross-correlation between material properties. %
In addition, the genetic algorithm has a high computational cost \cite{rylander2001computational}. %

An alternative Bayesian approach to this problem is to simulate samples from the posterior distribution using a Markov-Chain Monte Carlo (MCMC) procedure. %
This approach allows for an observation of the true shape of the posterior, as well as to simulate the multi-variate distributional behaviour of the parameters. %
Previously, using MCMC would have been impractical, as each iteration requires solving the dispersion curves given the current estimate of the material properties, which has a high computational expense. %
However, with the faster computation of dispersion-curve solutions (as discussed above), it is now much more feasible to apply such a procedure. %
In this work, the Legendre polynomial expansion approach \cite{Lefebvre2001} has been used, though many other options are available. %
A key argument for this choice is because, using some numerical manipulations, it was possible to increase the speed of this calculation. %
Adding further, the Legendre polynomial approach directly use the elastic constants, and so inference on these is more direct, and this allows for more efficient sampling in the MCMC process. %
The primary increase in computational efficiency is a result of the problem form being that of an eigenvalue problem in which the eigenvalues are the negative of the phase velocity squared; which then allows calculation of only the first modes of interest by using the power iteration method. %

In this work, propagation of Lamb waves in a glass-fibre-reinforced-polymer (GFRP) plate are measured and dispersion curve data are returned at various propagation angles. %
These dispersion curve data are then fed into the MCMC procedure and the posterior distributions of the material properties are analysed. %
In this initial work, the dispersion-curve solution method used does not include material damping. %
As the experimental observations of the dispersion-curves are normalised, material damping is not likely to affect simulation over the non-complex elastic constants which are used in the model. %
The limitations of this is that modelling of attenuation affects is restricted to not including viscoelastic effects. %
For brevity, the work in this paper considers only the real wavenumber modelling and observations. %

The next section of this paper begins with explanation of how to obtain dispersion curve solutions, including: numerical solutions, experimental observations, and the Legendre polynomial expansion (LPE) approach for orthotropic materials. %
Using the LPE approach, a brief sensitivity investigation is included to discuss the effects of each material parameter on the dispersion curve solutions. %
Section 3 details the experimental method for returning observations of the dispersion curves for the plate. %
The remainder of Section 3 then details how to estimate the elastic constants given observations, and how this is extended to use the MCMC approach to simulate sampling from the posterior. %
The paper then finishes by presenting and discussing the results of the procedure, along with a discussion of suggested future work that the authors intend to pursue. %

%%%%%%%%%%%%%%%%%%%%%%%%%%%%%%%%%%%%%%%%%%
\section{Lamb Wave Propagation in Plates}

In order to better apply guided waves for SHM and NDE strategies, prior knowledge of their behaviour is essential. %
This section aims to introduce the physics of guided waves, the concept of dispersion curves, and important characteristics which are prevalent in this work. %

\subsection{Physics of Lamb waves}
\label{sec:lamb_phys}

Elastic waves in orthotropic, inhomogeneous media are described by the elastodynamic equation \cite{Achenbach1973},
\begin{equation}
   \partial_l(S_{klmn}\partial_n w_{m}) = \rho\ddot{u}_k \qquad (k,l,m,n = 1,3)
   \label{eq:elastodynamic}
\end{equation}
where $S$ is the four-index stiffness tensor, $\rho$ is the material density, $u$ is the displacement field for which the double dot represents double differentiation with respect to time. %
In bounded media, these waves will show as Lamb waves, which in isotropic elastic media will exhibit two distinct modes: symmetric and antisymmetric. %
For anisotropic or composite media, there also exists shear-horizontal modes as a solution to \Cref{eq:elastodynamic}. %

When modelling guided waves in isotropic materials, the solutions to the two fundamental equations, derived from \Cref{eq:elastodynamic}, given the relationship between frequency $\omega$ and wavenumber $k$. %
The known frequency and wavenumber can then be used to determine the phase and group velocity, $c_p$ and $c_g$ respectively, using,
\begin{equation}
    c_p = \frac{\omega}{k}, \qquad c_g = \frac{d\omega}{dk}
    \label{eq:wave_relations}
\end{equation}
As the velocity of the wave is a function of the frequency, the waves are \emph{dispersive} and plots of the relationship between frequency and wavenumber/velocity are called \emph{dispersion curves}. %

\subsection{Dispersion Curve Solutions for Orthotropic Media}
\label{sec:orth_disp_solve}

Solutions of the dispersion curves for anisotropic media are more demanding. %
Here, the work of Lefebvre \cite{Lefebvre2001} is followed to formulate a computationally-efficient method of solving for these curves. %
This method has been validated in the works of Othmani \cite{othmani2016investigation,othmani2018influences}, as well as its improved computational efficiency demonstrated. %
This method uses a Legendre polynomial expansion to form an eigenvalue problem, utilising the orthonormal basis set for expansion of the field quantities. %
For orthotropic materials, the generalised Hooke's law can be rewritten as,
\begin{equation}
    \begin{Bmatrix}
        \sigma_{11} \\ \sigma_{22} \\ \sigma_{33} \\ \sigma_{23} \\ \sigma_{13} \\ \sigma_{12} 
    \end{Bmatrix}
    =
    \begin{bmatrix}
        C_{11} & C_{12} & C_{13} &    0    &    0    & 0  \\
               & C_{22} & C_{23} &    0    &    0    & 0  \\
               &        & C_{33} &    0    &    0    & 0  \\
               &        &        & C_{44} &    0    & 0  \\
               &        &        &        & C_{55} &  0 \\
               &        &        &        &        & C_{66} \\
    \end{bmatrix}
    \begin{Bmatrix}
        \varepsilon_{11} \\ \varepsilon_{22} \\ \varepsilon_{33} \\ 2\varepsilon_{23} \\ 2\varepsilon_{13} \\ 2\varepsilon_{12} 
    \end{Bmatrix}
\end{equation}
where $\sigma_{ij}$ is the stress, $\varepsilon_{ij}$ is the strain, and $C_{ij}$ is the elastic constants. %
The elastic constant tensor $C$ is the inverse of the stiffness matrix $S$, and the elements are defined using Voigt notation.  %
The relationship between the strain and displacement can be expressed as,
\begin{equation}
    % \begin{split}
        % \varepsilon_{ii} &= \frac{\partial u_i}{\partial x_i} \\
        \varepsilon_{ij} = \frac{1}{2}\left(\frac{\partial u_i}{\partial x_j} + \frac{\partial u_j}{\partial x_i} \right)
    % \end{split}
    \label{eq:strain_ortho}
\end{equation}
The boundary conditions of zero stresses on the surface can be applied by introducing a rectangular window function $\pi_{h}(x_3)$,
\begin{equation}
    \pi_{h}(x_3) = 
    \begin{cases}
        1 & 0 \leq x_3 \leq h \\
        0 & \text{otherwise}
    \end{cases}
\end{equation}
the above-mentioned boundary conditions are automatically incorporated in the constitutive relations, and by substituting in the relationship for the strain (\Cref{eq:strain_ortho}), and transforming the spatial coordinates into dimensionless form $q_{\alpha}$,
\begin{equation}
    q_{\alpha} = kx_{\alpha}, \qquad (\alpha=1,3)
\end{equation}
The constitutive relations are then,
\begin{equation}
    \sigma_{ij} = \left(C_{ijkl}\frac{\partial u_l}{\partial q_k}\right) k\pi_{h}(q_3)
    \label{eq:stress_const}
\end{equation}
For a wave propagating in the $x_1$ direction, the displacement components are assumed to be of the form
\begin{equation}
    u_i(q_1,q_2,q_3,t) = U_i(q_3)e^{\textrm{i}(q_1-\omega t)}
    \label{eq:displ_comp}
\end{equation}
where $U_i(q_3)$ represent the magnitudes of the fields in the $x_i$ direction, and the non-italic i is the imaginary unit. %
Substituting \cref{eq:stress_const,eq:displ_comp} into \Cref{eq:elastodynamic} gives,
\begin{subequations}
    \begin{align}
        \begin{split}
            -\frac{\omega^2}{k^2} U_1 = &-U_1\frac{C_{11}}{\rho} + iU'_3\left(\frac{C_{13} + C_{55}}{\rho}\right) + U''_1\frac{C_{55}}{\rho} \\
            &+ iU_3\frac{C_{55}}{\rho}\left(\delta(q_3)-\delta(q_3=kh)\right) + U'_1\frac{C_{55}}{\rho}\left(\delta(q_3)-\delta(q_3=kh)\right)
        \end{split} \label{eq:orth_vib_eq_a} \\
        \begin{split}
            -\frac{\omega^2}{k^2} U_2 = &-U_2\frac{C_{66}}{\rho} + U_2''\frac{C_{44}}{\rho} + U_2'\frac{C_{44}}{\rho}\left(\delta(q_3)-\delta(q_3=kh)\right)
        \end{split} \label{eq:orth_vib_eq_b} \\
        \begin{split}
            -\frac{\omega^2}{k^2} U_3 = &-U_3\frac{C_{55}}{\rho} + iU'_1\left(\frac{C_{31} + C_{55}}{\rho}\right) + U''_3\frac{C_{33}}{\rho} \\
            &+ iU_1\frac{C_{13}}{\rho}\left(\delta(q_3)-\delta(q_3=kh)\right) + U'_3\frac{C_{55}}{\rho}\left(\delta(q_3)-\delta(q_3=kh)\right) \label{eq:orth_vib_eq_c}
        \end{split}
    \end{align}
    \label{eq:orth_vib_eq}%
\end{subequations}
where the superscript $(\cdot)'$ refers to the partial derivative with respect to $q_3$. %
It is clear to see that \Cref{eq:orth_vib_eq_b} is independent of the other two equations; in fact, \Cref{eq:orth_vib_eq_b} represents the vibration of the SH waves in orthotropic plates and \cref{eq:orth_vib_eq_a,eq:orth_vib_eq_c} control propagation of Lamb wave modes. %

In order to solve the decoupled wave equations, the Legendre polynomial method expands $U_i(x_3)$ into an orthonormal polynomial basis \cite{Lefebvre2001,othmani2016investigation,othmani2018influences},
\begin{equation}
    U_i(q_3) = \sum_{m=0}^{\infty}p_m^iQ_m(q_3), \qquad i=1,2,3
    \label{eq:ortho_poly}
\end{equation}
where $p_m^i$ is the expansion coefficient and,
\begin{equation}
    Q_m(q_3) = \sqrt{\frac{2m+1}{kh}}P_m(\frac{q_3}{kh}-1)
    \label{eq:qm_expand}
\end{equation}
where $P_m(x)$ is the Legendre polynomial of order $m$. %
Theoretically, $m$ runs from 0 to $\infty$; however, in practice, the summation over polynomials in \Cref{eq:ortho_poly} can be halted at some finite value of $m=M$, when higher-order terms become negligible. %

To retrieve the final equations for solution, one substitutes \cref{eq:ortho_poly,eq:qm_expand} into \Cref{eq:orth_vib_eq}, multiplies by $Q_j^*(q_3)$ and integrates over $q_3$ from 0 to $kh$, giving,
\begin{subequations}
    \begin{align}
        \frac{\omega^2}{k^2}p_m^1 &= -M_{jm}^{-1}\left[A_{11}^{jm}p_m^1 + A_{13}^{jm}p_m^3\right] \\
        \frac{\omega^2}{k^2}p_m^2 &= -M_{jm}^{-1}\left[A_{22}^{jm}p_m^2\right] \\
        \frac{\omega^2}{k^2}p_m^3 &= -M_{jm}^{-1}\left[A_{31}^{jm}p_m^1 + A_{33}^{jm}p_m^3\right] 
    \end{align}
    \label{eq:orth_disp_final_eq}%
\end{subequations}
with $j$ and $m$ running from 0 to $M$, and $(\cdot)^*$ indicates the complex conjugate. %
The definitions of the matrix elements are shown in \Cref{app:orth_disp_mat}. %

By separating out \Cref{eq:orth_disp_final_eq} into only the coupled Lamb wave modes and the decoupled SH wave mode, the final solution can be arranged as an eigenvalue problem,
\begin{subequations}
    \begin{align}
        \begin{bmatrix}
            A_{11}^{jm} & A_{13}^{jm} \\
            A_{31}^{jm} & A_{33}^{jm}
        \end{bmatrix}
        \begin{bmatrix}
            p_m^1 \\ p_m^3
        \end{bmatrix}
        &= -\frac{\omega^2}{k^2} M_{jm}
        \begin{bmatrix}
            p_m^1 \\ p_m^3
        \end{bmatrix} \\
         \left[A_{22}^{jm}\right] p_m^2
        &= -\frac{\omega^2}{k^2} M_{jm} p_m^2
    \end{align}%
    \label{eq:eig_problem_form}
\end{subequations}
with eigenvalues $-c_p^2$ and corresponding eigenvectors $\{p_m^1 p_m^3\}^{\top}$. %
Here, $3(M+1)$ eigenmodes are generated at order $M$ of the expansion. %
The only solutions to be accepted are those eigenmodes for which convergence is obtained as $M$ is increased \cite{Lefebvre2001,othmani2016investigation,othmani2018influences}. %

These equations are not individually inferrable for respective wave modes, instead the matrix elements must be determined to form the eigenvalue problem in \Cref{eq:eig_problem_form}, the solutions of which provide dispersion information. %
Solutions to all available modes must be determined simultaneously using the eigenvalue solution, where the number of modes available is $2(M+1)$ for the Lamb wave modes, and $M+1$ for the SH wave modes. %

Previously, this method has been implemented using a symbolic-programming approach; however, as the expansion forms a series of polynomials, a programmatic approach has been developed here to reduce computation time. %
This strategy was in fact the first of many numerical tactics employed to reduce computation cost in order to make the method applicable to a probabilistic sampling procedure such as MCMC. %
More details on all the manipulations developed can be found in \Cref{app:num_tricks}. %

A particularly noteworthy aspect of the numerical manipulations implemented is the use of the power iteration method to determine the eigenvalue solutions. %
The power method allows one to determine, in sequence, the eigenvalues from largest to smallest, at a lower computational cost \cite{Mises1929,tufts1986simple}. %
Therefore, in order to maintain a computational efficiency, only the first two wave modes ($A_0$ and $S_0$) are included here. %

\subsection{Prior Exploration of Model}
\label{sec:disp_curve_sens}

\begin{figure}[h!]
	\begin{adjustwidth}{-\extralength}{0cm}
    \centering
    \begin{subfigure}{6cm}
        \includegraphics[width=6cm]{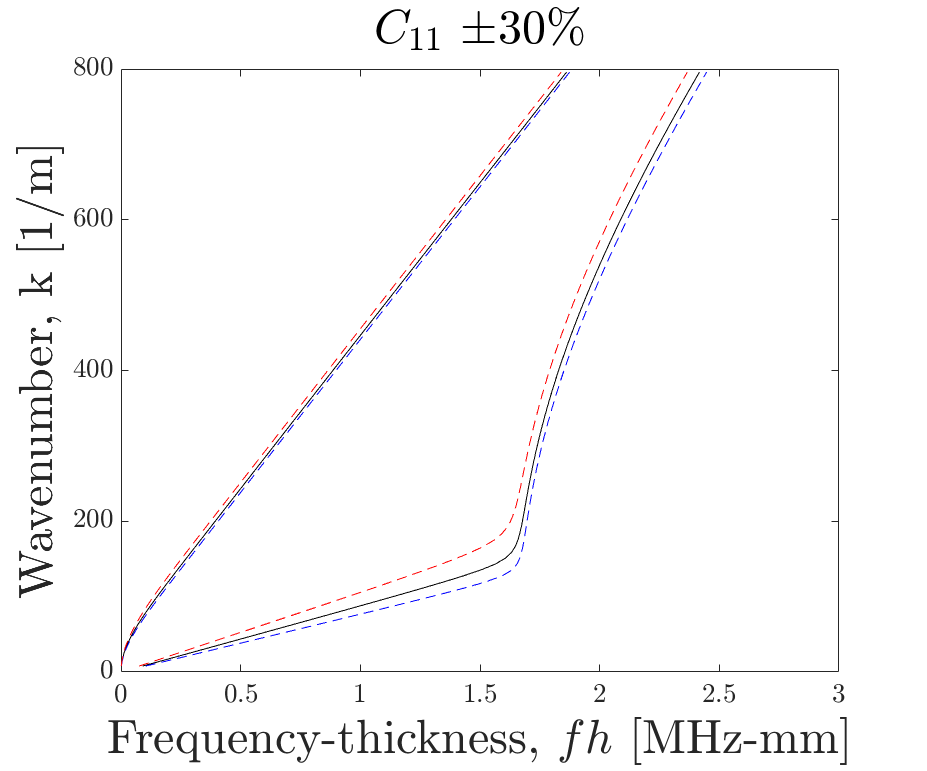}
        \caption{}
    \end{subfigure}
    \begin{subfigure}{6cm}
        \includegraphics[width=6cm]{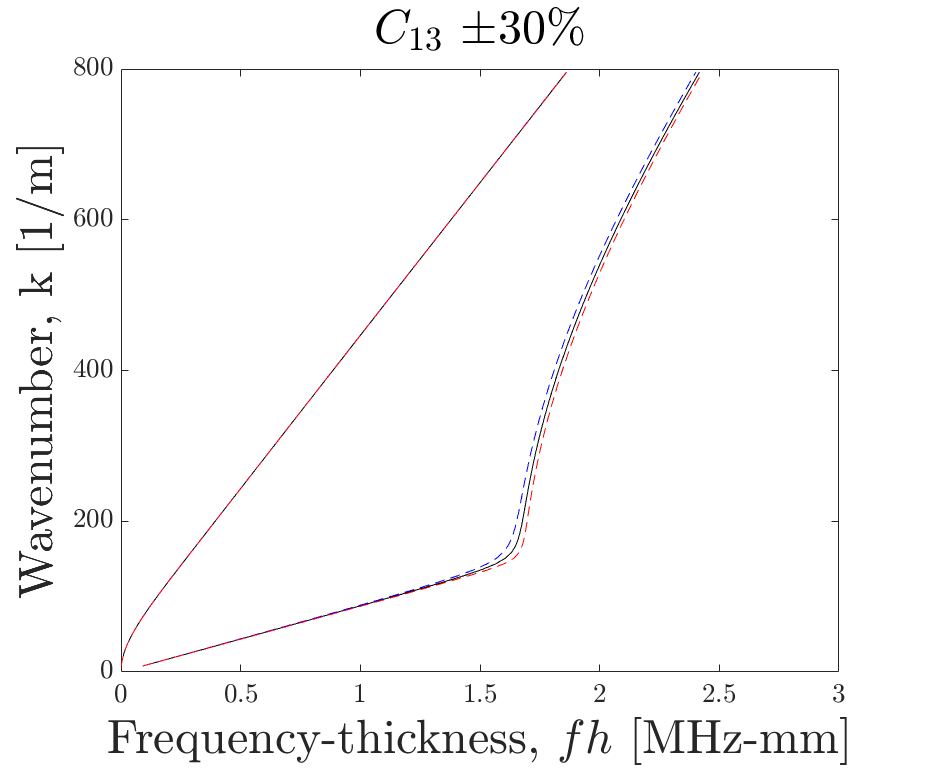}
        \caption{}
    \end{subfigure}
    \begin{subfigure}{6cm}
        \includegraphics[width=6cm]{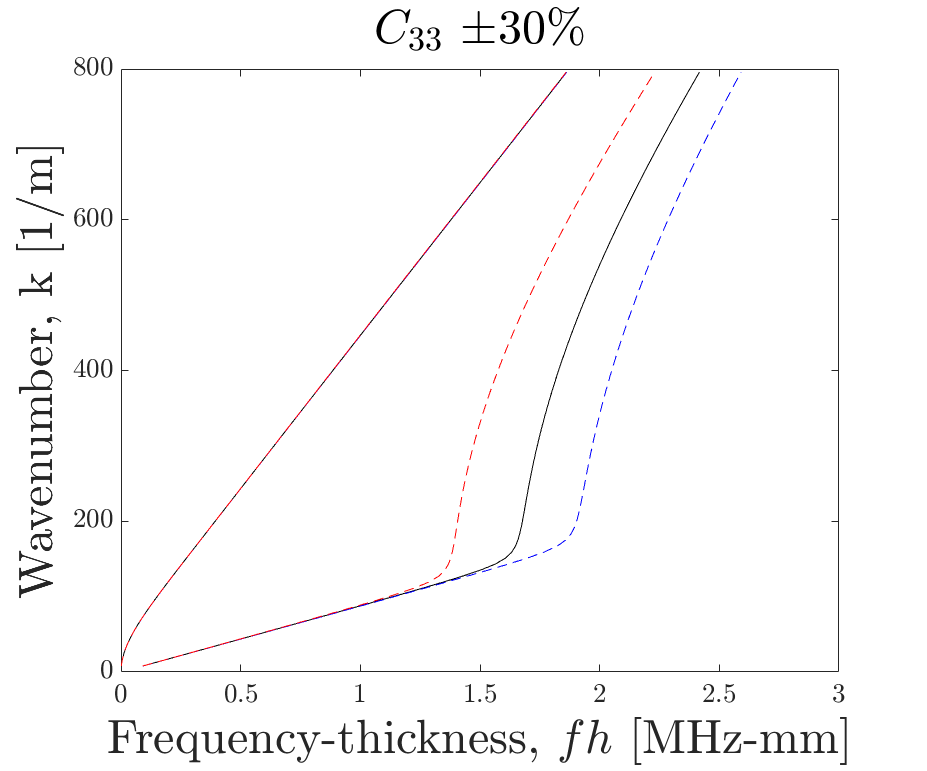}
        \caption{}
    \end{subfigure}
    \begin{subfigure}{6cm}
        \includegraphics[width=6cm]{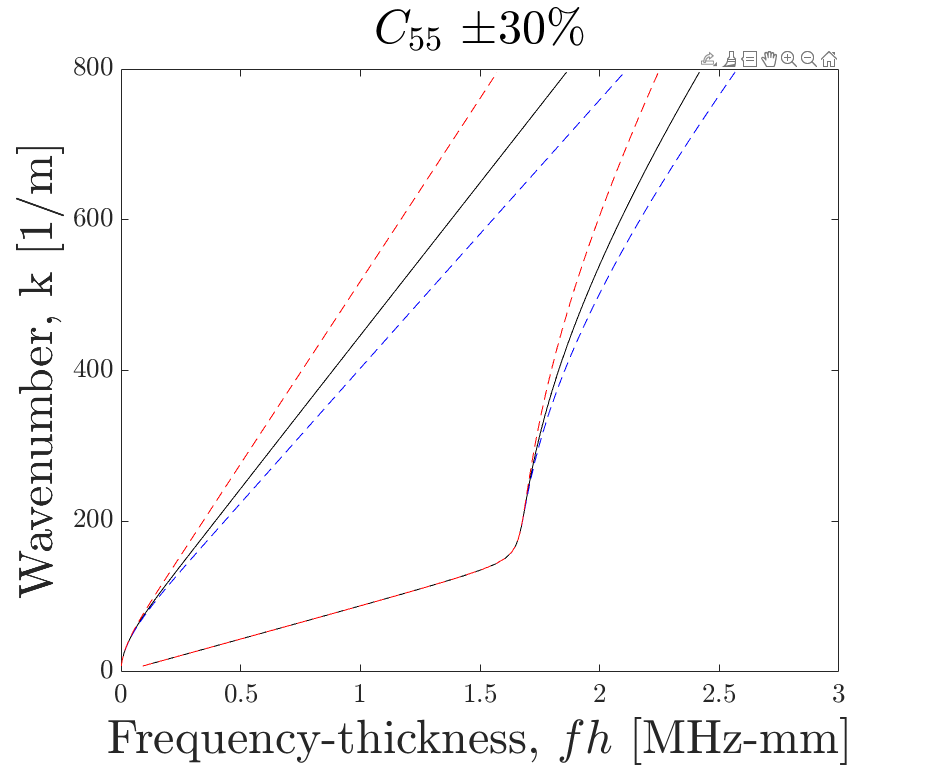}
        \caption{}
    \end{subfigure}
    \begin{subfigure}{6cm}
        \includegraphics[width=6cm]{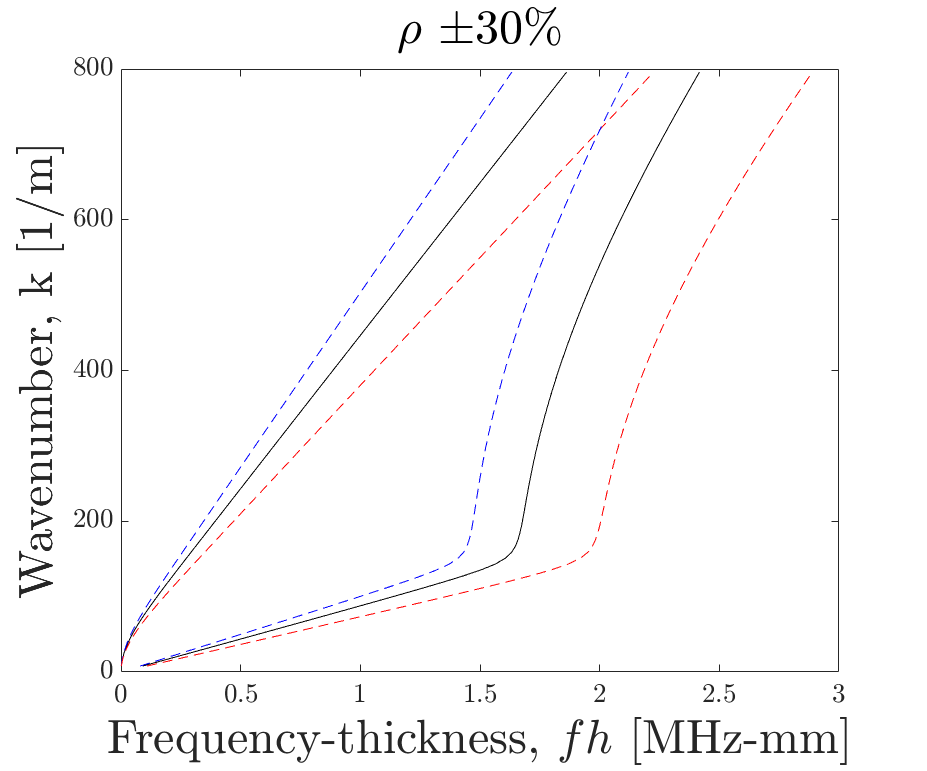}
        \caption{}
    \end{subfigure}
	\end{adjustwidth}
    \caption{Sensitivity analysis of material constants on dispersion curves of fundamental Lamb wave modes, for parameters (a) $C_{11}$, (b) $C_{13}$, (c) $C_{33}$, (d) $C_{55}$, and (e) $\rho$. With initial values $C_{11}$ = 160\si{GPa}, $C_{13}$ = 6.5\si{GPa}, $C_{33}$ = 14\si{GPa}, $C_{55}$ = 7\si{GPa}, and $\rho$ = 1200\si{kg m^{-3}}. For all figures, the black line shows the solution with initial values, the blue line shows the solution with a change of +30\% and the red line a change of -30\%.}
    \label{fig:disp_param_effects}
\end{figure}

Before progressing to the procedure to identify the material properties governing \Cref{eq:orth_disp_final_eq}, it is useful to explore the effects on the dispersion curves of changes in the material properties. %
\Cref{fig:disp_param_effects} shows how the curves for the fundamental Lamb wave modes are affected by changes in the elastic constants and density of the orthotropic model. %
The initial values of each were chosen based on those used in \cite{othmani2016investigation}. %

Changes in all elastic constants appear to be consistently stronger in the solutions for the $S_0$ mode. %
The constants $C_{13}$ and $C_{33}$ appear to have very little effect on the dispersion curve for the $A_0$ mode, but have significant effects on the $S_0$ mode. %

An interesting observation is on the effect of $C_{33}$ and $C_{55}$ on the $S_0$ curve; no significant changes appear until after the `elbow' in the curve -- the sharpness of which is unique to more complex models and does not appear in isotropic dispersion curves. %
For $C_{55}$ and $\rho$, changes appear to be stronger at higher frequencies, whereas the changes appear more consistent across the frequency range for $C_{11}$, $C_{13}$ and $C_{33}$. %

%%%%%%%%%%%%%%%%%%%%%%%%%%%%%%%%%%%%%%%%%%
\section{Material Indentification Procedure}
\label{sec:methodology}

\subsection{Measuring observations of dispersion curves}
\label{sec:meas_wk}

The first stage of the process here is to determine a set of measured values on the dispersion curve $\{\hat{\omega},\hat{k}\}$ of the plate in question. %
Dispersion curves can also be determined from arbitrary plates by the use of a \emph{two-dimensional Fourier transform} (2DFT); this is done by recording the surface displacement of a Lamb wave, spatially sampled along its propagation path, to generate the time-distance [$\var{t-x}$] space. %
Passing this through a 2DFT then provides a transformation to the frequency-wavenumber [$\omega\var{-k}$] space \cite{Alleyne1991}. %
The surface displacement of a wave at regularly-spaced intervals is measured to form time-distance [$\var{t-x}$] data. %
The signals at each spatial location are then normalised and the matrix passed through a 2DFT to retrieve the frequency-wavenumber [$\omega\var{-k}$] data. %

\begin{table}[h!]
    \centering
    \begin{tabular}{|l|l|}
        Plate & GFRP \\
        % \hline
        Plate dimensions & \qtyproduct{800 x 700 x 16}{mm}\\
        % Layup & $[90/0/90]_s$, Epoxy matrix & Unidirectional\\
        PZT Location & \qtyproduct{400 x 235}{mm}\\
        Actuation signal & 500\si{kHz} chirp\\
        Signal record length & 8\si{\milli s}\\
        Sampling frequency & 1.024\si{\mega Hz}\\
        Spatial sampling step size & 1.8\si{mm}\\
        No. averages & 100 \\
    \end{tabular}
    \caption{Details of experimental setup used to acquire Lamb wave signal data. The datum which is used for the PZT location details is the bottom left corner of the plate, as indicated in \Cref{fig:exp_setup}.}
    \label{tab:exp_details}
\end{table}

\begin{figure}[H]
	\begin{adjustwidth}{-\extralength}{0cm}
    \centering
    \includegraphics[width=16.5cm]{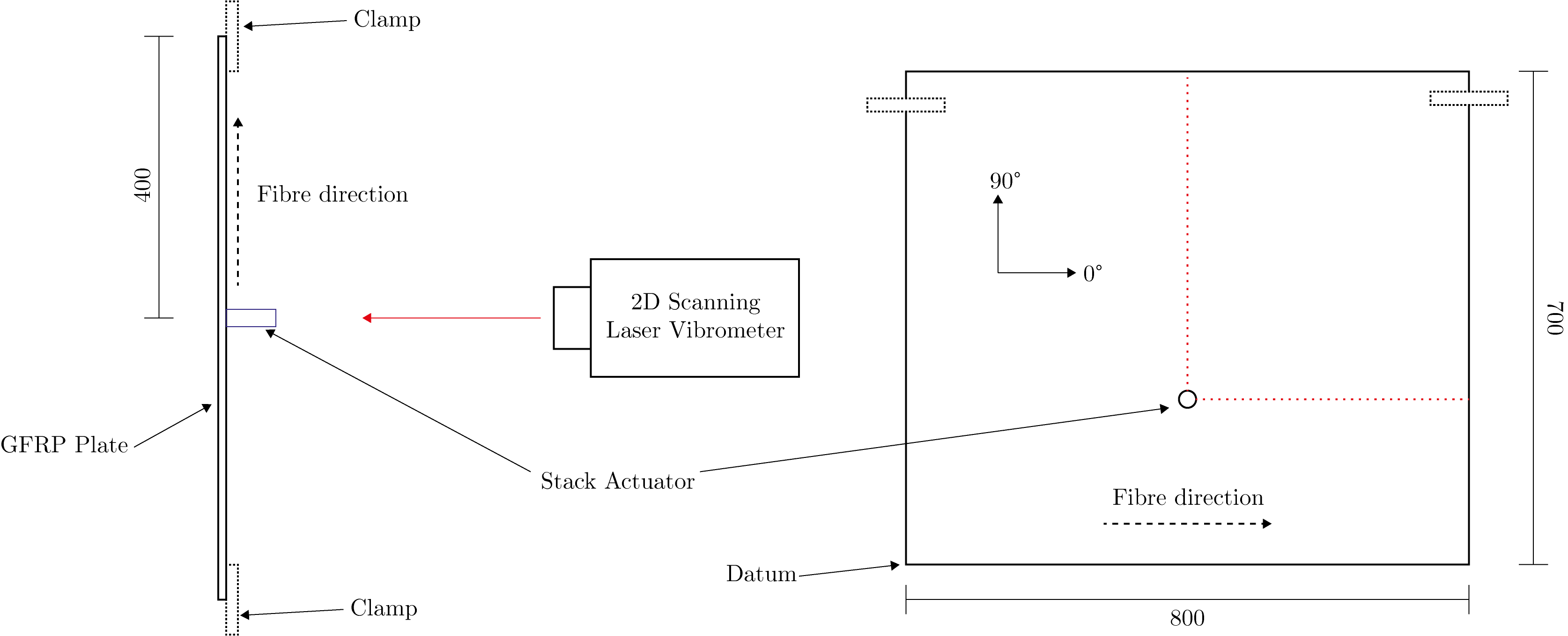}
	\end{adjustwidth}
    \caption{Diagram of the experimental setup and location of stack actuator on the \qtyproduct{800 x 700}{mm} GFRP plate. The left diagram shows a top-down view, and the right a front view. The orientation of the fibres and the coordinate system used for naming conventions is also shown. The red dotted line shows the lines along which the laser scanner recorded surface displacement.}
    \label{fig:exp_setup}
\end{figure}

Lamb waves were initiated in a glass-fibre reinforced-polymer (GFRP) plate by excitation of a 20mm diameter piezo-electric transducer (PZT) stack actuator (Physik Instrumente P-016.20P), on the surface of the plate, as shown in \Cref{fig:exp_setup}. %
The PZT was actuated with a chirp signal of length 1ms and upper frequency of 500kHz, allowing for broadband excitation. %
A Polytec PSV-400 scanning-laser vibrometer was used to measure the out-of-plane surface displacement of the induced wave-packets along a single propagation path, where the recording state was synchronised with the start of the excitation signal. %
Retro-reflective tape, 0.5mm thick, was placed along these propagation paths to improve the signal-to-noise ratio. %
Information on the material properties for the plate are shown in \Cref{tab:gfrp_plate_props}, however, it is important to note that these are for model validation purposes and are not fed into the methodology. %

\bgroup
\def\arraystretch{1.5}
\begin{table}[h!]
    \centering
    \begin{tabular}{l||l|l|l|l|l|l||l|l|l|l}
        & \multicolumn{6}{c||}{Provided Material Properties} & \multicolumn{4}{c}{Calculated ECs} \\
        \hline
        Property & $E_{11}$ & $E_{22}$ & $G_{12}$ & $\nu_{12}$ & $\nu_{21}$ & $\rho$ & $C_{11}$ & $C_{13}$ & $C_{33}$ & $C_{55}$ \\
        \hline
        Value (GPa) & 24.5 & 14.6 & 8.2 & -0.46 & -0.28 & 1200 & 28.1 & 7.8 & 16.7 & 8.2 \\
    \end{tabular}
    \caption{Given material properties, and calculated elastic constants, of the GFRP coupon.}
    \label{tab:gfrp_plate_props}
\end{table}
\egroup

A PZT stack actuator was used, as opposed to a disc, in order to improve the signal-to-noise ratio; as the plate used is relatively thick, and the epoxy matrix causes rapid attenuation of the waves. %
The location of the PZT stack actuator was chosen in order to maximise propagation distance before reflection due to boundaries. %
It was placed ~1/3 of the distance in the vertical direction, which would result in reflections from the upper and lower boundaries imposing on the propagation wavefield at the same time. %
It would initially be preferable to do the same thing for the horizontal direction also, however, the propagation velocity in this direction will be larger as it is in the direction of the fibres, and so may impose on the vertical propagation wavefield before the wave reflection at the boundary. %

\begin{figure}[H]
    \centering
    \includegraphics[width=0.55\textwidth]{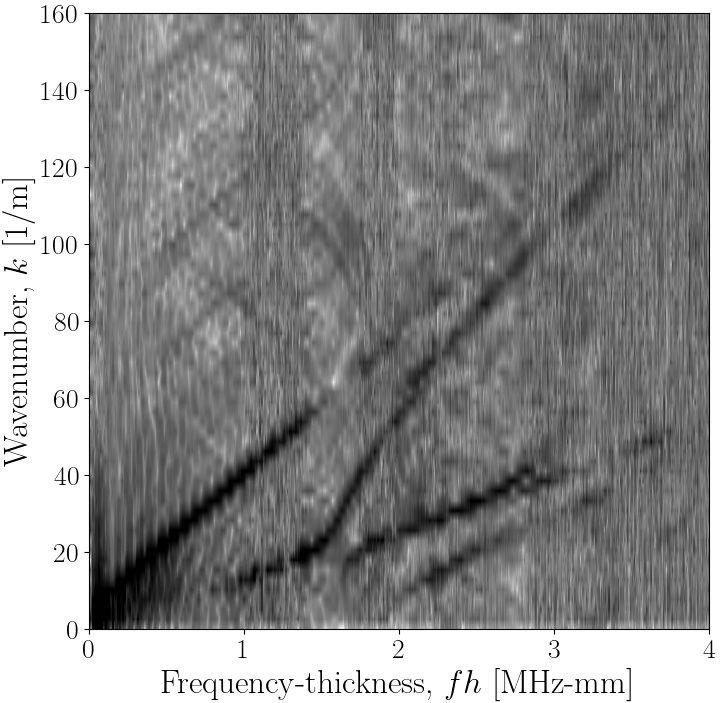}
    \caption{Normalised dispersion curve image data for propagation angle of 0\si{\degree}.}
    \label{fig:disp_im_GFRP}
\end{figure}

Data were recorded along two propagation path directions; 0\si{\degree} and 90\si{\degree}, these path directions are also shown in \Cref{fig:exp_setup}. %
Specific details of the experimental setup are shown in \Cref{tab:exp_details}, including plate dimensions and acquisition parameters. %
The [$\var{t-x}$] data were then passed through a 2DFT to form dispersion-curve images at each angle; the results for 0\si{\degree} are shown in \Cref{fig:disp_im_GFRP}. %

As the natural frequency of the stack actuator is within the frequency range of the dispersion curves of interest, there will be disparities in energy content along this frequency range. %
Therefore, to improve contrast of the image, resulting in a more equal distribution of observations along the frequency axis, the image data were normalised with respect to the energy content. %
This was done by dividing the elements of each frequency bin vector were divided by the sum of the frequency vector content. %
Where $U$ is the image data, $\tilde{U}$ is the normalised dispersion-curve image data, $n_k$ is the length of the frequency vector, $k$ represents the wavenumber index and $f$ the frequency index,
\begin{equation}
    \tilde{U}_f = U_f/\left(\sum_{i=1}^{n_k}U_{f,k}\right)
\end{equation}

The dispersion-curve image data for the GFRP plate in \Cref{fig:disp_im_GFRP} shows strongly both the $A_0$ and $S_0$ modes, as well as some information present on the $S_1$ and $S_2$ modes. %
Using the ridge-picking algorithm for this data, as described in \cite{Dobie2011}, only the $A_0$ and $S_0$ modes are considered. %
From the experimental setup, the upper limit of the frequency-thickness bandwidth is 8.192 \si{MHz-mm}. %
However, \Cref{sec:disp_curve_sens} showed that the dispersion curve solutions are more sensitive to changes in the material properties at higher frequencies, therefore, the frequency-thickness bandwidth of 4.098 \si{MHz-mm} was chosen to include all information available on the $A_0$ and $S_0$ modes. %

\subsection{Estimating Elastic Constants}

Consider the concept of an elementary system-identification procedure to estimate a set of parameters given a set of $n$ observations, %
\begin{equation}
	\mathbf{y} = \{y_1,y_2,...,y_n\}
\end{equation}
When using a probabilistic approach to determining elastic constants, one must form a definition of the \emph{likelihood} of a set of constants, given some observed data. %
Here, the likelihood is defined based on the choice of noise model, i.e.\ Gaussian. %
This likelihood could be used to retrieve a maximum likelihood estimate \cite{lecam1990maximum}, which is a popular and asymptotically-optimal statistical approach to fitting model parameters using data \cite{lehmann2006theory}. %
Assuming the model is of the form,
\begin{equation}
	y_i = f(r) = r+ \varepsilon_i
\end{equation}
where $r$ is the mean at point $n$, and $\varepsilon_i$ is a white Gaussian noise process. %
The observations are distributed as $y\sim\mathcal{N}(r,\sigma^2)$. %
The likelihood is then defined as,
\begin{equation}
    L(\mathbf{y}|r) = \prod_{i=1}^n \frac{1}{\sqrt{2\pi\sigma^2}} \exp\left(-\frac{1}{2}\frac{(y_i-r)^2}{\sigma^2}\right)
\end{equation}
For determining the likelihood of some model, the mean can be replaced with a function of the input dimension $\mathbf{x}$ of the observations, and some parameters $\Theta$,
\begin{equation}
    r = f(x_i,\Theta)
\end{equation}
so the likelihood becomes,
\begin{equation}
    L(\mathbf{y}|\Theta) = \prod_{i=1}^n \frac{1}{\sqrt{2\pi\sigma^2}} \exp\left(-\frac{\left(y_i-f(x_i,\Theta)\right)^2}{2\sigma^2}\right)
\end{equation}

In \Cref{sec:meas_wk}, it was noted that there is a much lower relative resolution in the wavenumber dimension of the dispersion image and in the resulting selected points on the curve. %
This implies that the Gaussian white-noise distribution is mostly in $\omega$; thus, if one were to estimate the likelihood based on a model of $k(\omega,\Theta)$, the function would be of the form,
\begin{equation}
    y_i = f(r + \varepsilon_i)
\end{equation}
Therefore, the problem is formulated as based of a model of $\omega(k,\Theta)$. %
The observations are taken as the points on the dispersion curve,
\begin{equation}
    y_i = \{\hat{\omega}_i,\hat{k}_i\}
\end{equation}
where $\hat{\omega}_i$ and $\hat{k}_i$ are the measured values of frequency and wavenumber respectively, at point $i$. %
The set of observations is grouped into $m$ modes, individually represented by $\psi$; $\mathbf{y}=\{\mathbf{y}_{\psi_1}^{\top},\mathbf{y}_{\psi_2}^{\top},...,\mathbf{y}_{\psi_m}^{\top}\}$, where $\mathbf{y}_{\psi}=\{\hat{\omega}_{\psi},\hat{k}_{\psi}\}$. %
For example, in the case where only the fundamental modes are considered, $\psi_1$ and $\psi_2$ represent the $A_0$ and $S_0$ modes respectively. %
The likelihood is then defined as,
\begin{equation}
    L(\mathbf{y}|\Theta) = \prod_{j=1}^m \prod_{i=1}^n \frac{1}{\sqrt{2\pi\sigma^2}} \exp\left(-\frac{\left(\hat{\omega}_{i,\psi_j}-\omega_{\psi_j}(\hat{k}_{i,\psi_j},\Theta)\right)^2}{2\sigma^2}\right)
    \label{eq:disp_likeli}
\end{equation}
In this case, $\omega(\hat{k}_i,\Theta)$ is determined using the methods outlined in \Cref{sec:orth_disp_solve}. %
For the work presented here, only the first antisymmetric mode $A_0$ is considered, and so only solutions for that curve are returned. %
The parameters are defined as the elastic constants which enter into the equations for the Lamb wave modes in \cref{eq:orth_vib_eq_a,eq:orth_vib_eq_c},
\begin{equation}
    \Theta = \{C_{11},C_{13},C_{33},C_{55},\rho\}
\end{equation}
as $C_{31}=C_{13}$. %
Maximising $L(\mathbf{y}|\Theta)$ provides an estimate of the most likely elastic constants; however, it is also possible to retrieve information on their \emph{distribution}. %

\subsection{Estimating the posterior distributions}

The objective at this stage is to determine the distribution of the parameters which could plausibly define the dispersion curve. %
As the likelihood includes a noise variance term $\sigma$, the parameter vector is extended to include this, so that,
\begin{equation}
    \theta=\{\Theta,\sigma\} = \{C_{11},C_{13},C_{33},C_{55},\rho,\sigma\}
\end{equation}
The distribution of these parameters can be determined by identifying the posterior probability given a set of measured data, $p(\theta|\mathbf{y})$. %
However, this is not directly inferable, so a manipulation is done using Bayes rule,
\begin{equation}
    p(\theta|\mathbf{y}) = \frac{p(\mathbf{y}|\theta)p(\theta)}{\int p(\mathbf{y}|\theta)p(y)p(\theta)}
    \label{eq:bayes_rule}
\end{equation}
where $p(\mathbf{y}|\theta)$ is calculated using \Cref{eq:disp_likeli}, and $p(\theta)$ is the \emph{prior}, which can be defined using initial knowledge of the parameters. %
For $d$ parameters, assuming each parameter is independent, the prior is calculated as,
\begin{equation}
    p(\theta) = \prod_{i=1}^d p(\theta_i)
\end{equation}
Now, the problem is transferred, in that the normalisation term in the denominator is intractable. %
Instead, a procedure can be done to sample from the posterior with enough repetition that an estimate of the distribution over the parameters can be inferred. %
One such procedure is the Markov-Chain Monte Carlo (MCMC) method, where subsequent samples depend on assessing their probability with respect to the previous one. %
An outline of the derivation and procedure for MCMC is given in \cite{gamerman2006markov,gilks1995markov,barber2012bayesian}. %
In practice, for computational stability, the probabilities are calculated in the log space, so the marginal likelihood becomes,
\begin{equation}
    \log(\hat{p}(\theta|\mathbf{y})) = \log(p(\mathbf{y}|\theta)) + \log(p(\theta))
\end{equation}
where,
\begin{equation}
    \log(p(\theta)) = \sum_{i=1}^d \log(p(\theta_i))
\end{equation}

Now, consider how to define this problem for the application to dispersion curve material identification. %
The first step is to define the likelihood, which is done using \Cref{eq:disp_likeli},
\begin{equation}
    p(\mathbf{y}|\theta) = \prod_{j=1}^m \prod_{i=1}^n \frac{1}{\sqrt{2\pi\sigma^2}} \exp\left(-\frac{\left(\hat{\omega}_{i,\psi_j}-\omega_{\psi_j}(\hat{k}_{i,\psi_j},\Theta)\right)^2}{2\sigma^2}\right)
\end{equation}
which in the log space is,
\begin{equation}
    \log(p(\mathbf{y}|\theta)) = -mn\log(\sigma) - \frac{mn}{2}\log(2\pi) - \frac{1}{2} \sum_{j=1}^m \sum_{i=1}^n \frac{\left(\hat{\omega}_{i,\psi_j}-\omega_{\psi_j}(\hat{k}_{i,\psi_j},\Theta)\right)^2}{\sigma^2}
\end{equation}
During sampling using MCMC, the size of the random step taken for each parameter is important as too large a step will cause stall, and too small a step will require a large number of iterations. %
An improvement is made on the standard MCMC procedure, which incorporates Hamiltonian mechanics, to adapt the step size for an optimal simulation, and is so called Hamiltonian Monte Carlo (HMC) \cite{neal2011hamiltonian,betancourt2015hamiltonian}. %
For this work, the probabilistic programming language Stan \cite{stan2_28}, was used to perform the actual sampling procedure. %

\begin{table}[h!]
    \centering
    \begin{tabular}{l|l|ll}
    Parameter & Distribution                   & \multicolumn{2}{l}{Definition} \\
    \hline
    $C_{11}$  & $\textrm{Gamma}(\alpha,\beta)$ &      $\alpha=2$ & $\beta=0.02 $     \\
    $C_{13}$  & $\textrm{Gamma}(\alpha,\beta)$ &      $\alpha=1.5$ & $\beta=0.05 $     \\
    $C_{33}$  & $\textrm{Gamma}(\alpha,\beta)$ &      $\alpha=1.5$ & $\beta=0.05 $     \\
    $C_{55}$  & $\textrm{Gamma}(\alpha,\beta)$ &      $\alpha=1.5$ & $\beta=0.025 $     \\
    $\rho$    & $\mathcal{N}(\mu,\sigma_p)$      &      $\mu=1600$ & $\sigma_p=300 $     \\
    $\sigma$  & $\textrm{Gamma}(\alpha,\beta)$ &      $\alpha=2$ & $\beta=2\times 10^{-5} $     \\
    \end{tabular}
    \caption{Definitions of priors for parameters in $\theta$. }
    \label{tab:prior_defs}
\end{table}

Next, consider the definition of the priors, which can be done using reasonable knowledge of the material of application. %
As the prior is a combination of the individual probabilities of each parameter, prior belief on the distribution of these parameters can be used to define each $p(\theta_i)$. %
In this case, the density of the plate is supplied, but no other material properties were provided. %
Therefore, a tight prior can be given on $\rho$ and priors on the elastic constants are defined to capture reasonable values for the material. %
Here, a gamma distribution was chosen for the elastic constants as this provides a broad definition of the prior, which can be interpreted as embedding belief on the magnitude of the value, and enforces a positive-only value. %
As the prior for the density can be defined relatively tightly, this was defined using a normal distribution. %
The type and definitions of the priors used here are shown in \Cref{tab:prior_defs}. %

%%%%%%%%%%%%%%%%%%%%%%%%%%%%%%%%%%%%%%%%%%

\begin{figure}[H]
	% \begin{adjustwidth}{-\extralength}{0cm}
    \centering
    \includegraphics[width=13cm]{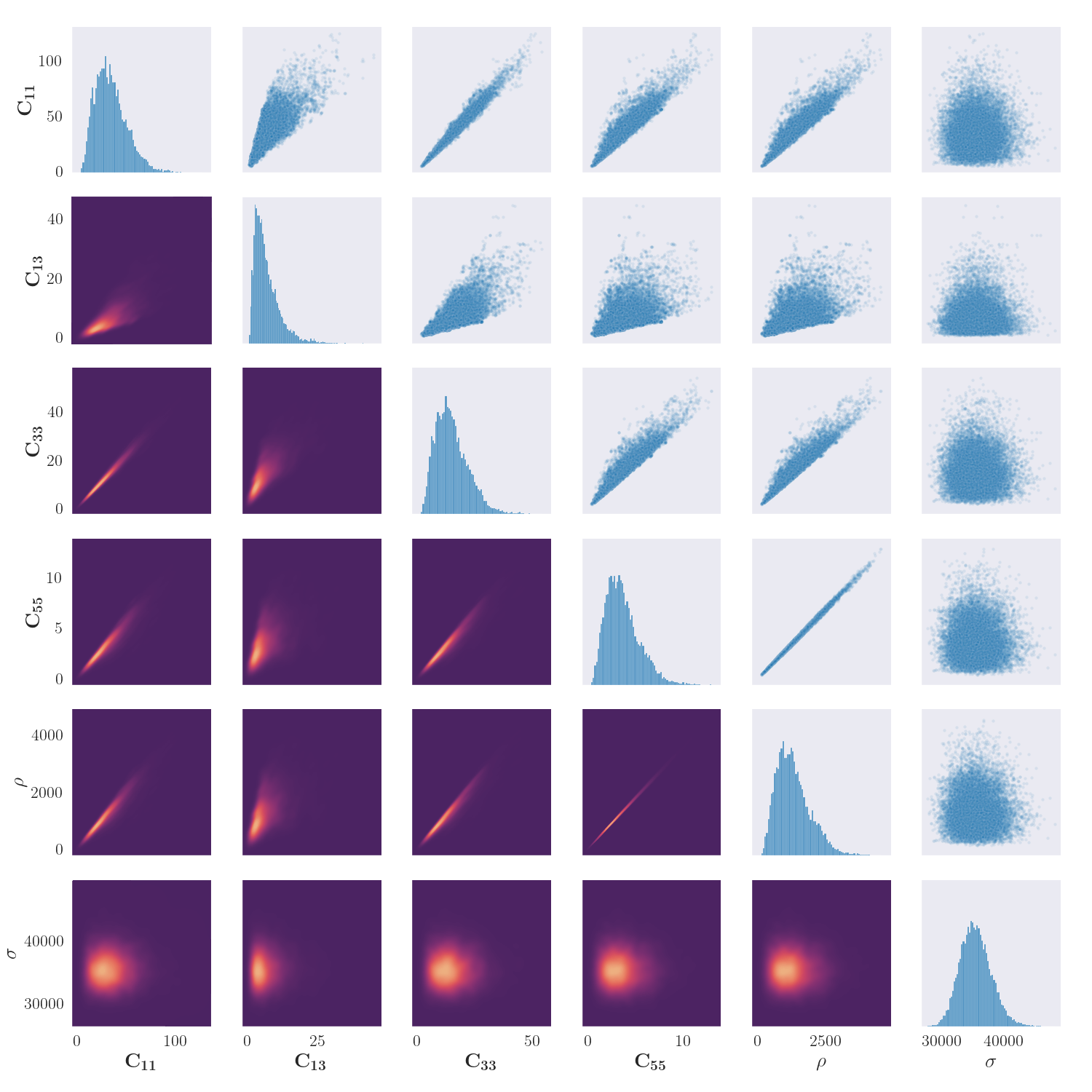}
	% \end{adjustwidth}
    \caption{Results of the parameter identification procedure applied to the blade coupon for the propagation angle of 0\si{\degree}. Figures along the diagonal show the histogram of the samples for each parameter. Figures in the upper right triangle show a scatter plot of correlation between two parameters. Figures in the lower left triangle show a bivariate kernel density estimate of the cross-correlation between parameters, where lighter colours represent a larger value of the density.}
    \label{fig:GFRP_correl_0_deg}
\end{figure}

\section{Results}
\label{sec:results}

In this section, samples from the posterior distributions of the parameters are shown in both univariate and bivariate distributions, and a kernel density estimate is used to estimate the probability density function of the bivariate distributions. %
Also shown, are samples of the dispersion curves drawn from the samples of the posterior distributions, overlaid onto the observed dispersion-curve image data taken from the two-dimensional Fourier transform of the measured surface displacement. %
% This section is split into three subsections; firstly, the results of the procedure applied to data where the wave propagates along the fibres are shown, followed by the results of the procedure applied to data where the wave propagates at 90\si{\degrees} to the fibres.
This section is split into three subsections; firstly, the simulated posterior distributions of the parameters are shown, followed by a simulated distribution of dispersion-curve data based on these parameters. %
Lastly, the first two statistical moments of the univariate samples are calculated; which are then used to display the estimated mean and variance of each parameter for each plate. %

\subsection{Posterior distribution of the material parameters}

\begin{figure}[H]
	% \begin{adjustwidth}{-\extralength}{0cm}
    \centering
    \includegraphics[width=13cm]{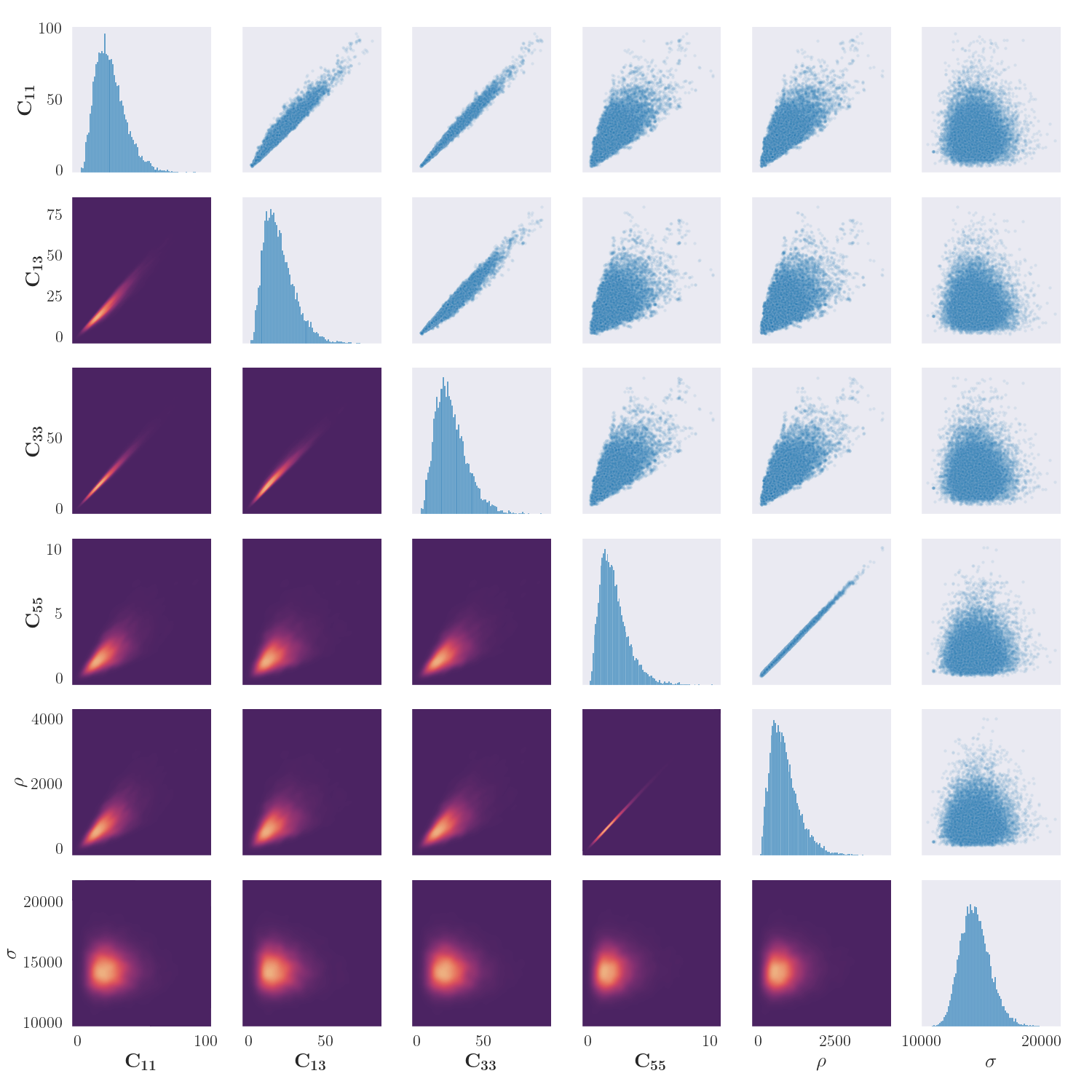}
	% \end{adjustwidth}
    \caption{Results of the parameter identification procedure applied to the blade coupon for the propagation angle of 90\si{\degree}. Figures along the diagonal show the histogram of the samples for each parameter. Figures in the upper right triangle show a scatter plot of correlation between two parameters. Figures in the lower left triangle show a bivariate kernel density estimate of the cross-correlation between parameters, where lighter colours represent a larger value of the density.}
    \label{fig:GFRP_correl_90_deg}
\end{figure}

The results of 20,000 accepted samples of the sampling procedure for propagation angles of 0\si{\degree} and 90\si{\degree} are shown in \Cref{fig:GFRP_correl_0_deg,fig:GFRP_correl_90_deg} respectively. %
The first observation that can be made is of the evidence of correlation between all material parameters, whereas the distribution of the noise parameter appears to converge to an independent distribution. %
This result is anticipated, as the elastic properties which form the stiffness matrix are described by a series of inseperable equations. %

\Cref{fig:GFRP_correl_0_deg} indicates the univariate and bivariate distributions for a propagation angle of 0\si{\degree}. %
There is an apparent `edge' on the scatter correlation plots between certain parameters, in particular between $C_{13}$ and all other elastic constants. %
As a condition of the solution to the dispersion curve equations is that $\lambda<0$, any solutions where this is not the case are rejected. %
The edge may indicate a region of forbidden parameter combinations which cannot exist, given a real elastic material. %

There is evidence of a particularly strong correlation between $C_{55}$ and $\rho$, which appears to be a linear relationship. %
This could be explained by comparison to the isotropic case; for an isotropic material their relationship can be defined as $C_{55} = \rho c_T^2$. %
For an orthotropic material, the transverse-wave velocity would remain the same when rotating around the axis in the direction of wave propagation. %
This property could be used to reduce the number of parameters, increasing performance of the simulation. %

One observation made in the results of the sample plots for the 90\si{\degree} propagation, in comparison to 0\si{\degree}, is the less apparent hard edge caused by the rejection parameter. %
This may indicate that the posterior space of valid elastic constants is less discontinuous when modelling Lamb-wave propagation through fibres. %
Another difference between the two sets of results is that there is a less strong correlation in the parameter pairs $C_{11}\var{-}C_{13}$ and $C_{13}\var{-}C_{33}$. %
This could be a result of the fibres no longer acting as a secondary wave guide, and instead shear forces through the fibres have more of an influence on wave propagation. %

\subsection{Distribution of dispersion curve models}

\begin{figure}[H]
	\begin{adjustwidth}{-\extralength}{0cm}
    \centering
    \begin{subfigure}{9cm}
        \centering
        \includegraphics[width=8.5cm]{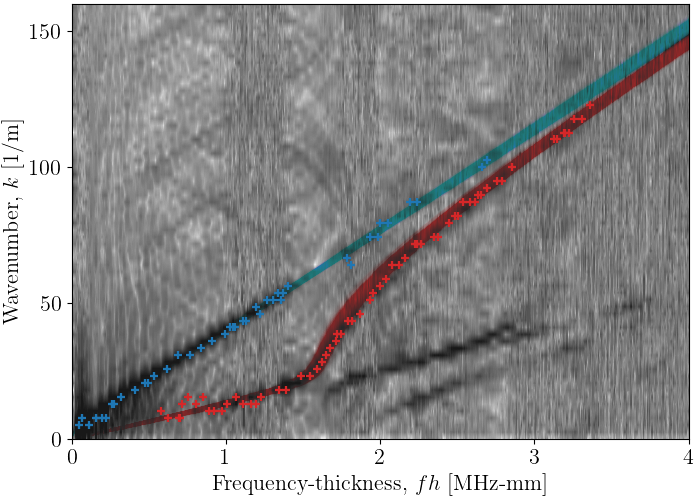}
        \caption{}
        \label{fig:GFRP_curve_distr_0_deg}
    \end{subfigure}
    \begin{subfigure}{9cm}
        \centering
        \includegraphics[width=8.5cm]{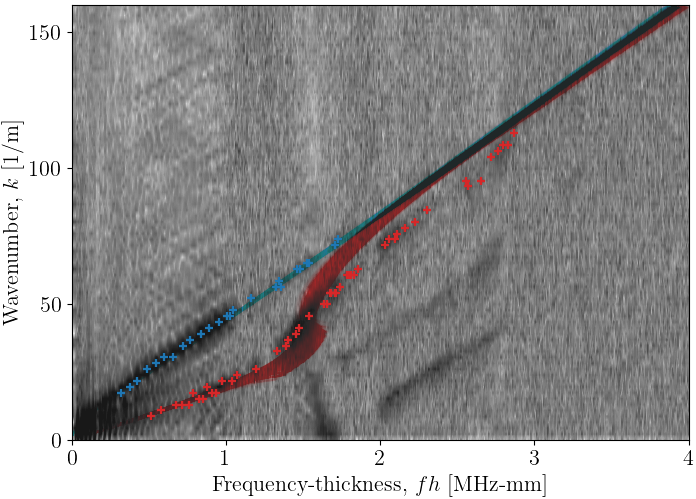}
        \caption{}
        \label{fig:GFRP_curve_distr_90_deg}
    \end{subfigure}
	\end{adjustwidth}
    \caption{Distribution of generated curves for each sample taken for propagation angles of (a) 0\si{\degree} and (b) 90\si{\degree}. The blue curves show the distribution of the $A_0$ mode and red curves show the distribution of the $S_0$ mode. The curves are overlaid on the image data taken from the 2DFT and the `+' markers indicate the points taken from the ridge-selection algorithm which were used in the procedure as $\{\hat{\omega},\hat{k}\}$.}
    \label{fig:GFRP_curve_distrs}
\end{figure}

Using the parameters at each sample point, a distribution of dispersion curves was generated, and is shown in \Cref{fig:GFRP_curve_distrs}, along with observations taken from the [$\omega\var{-k}$] image data. %
For the propagation angle of 0\si{\degree}, the darker areas of the image data, as well as the observation points, lie within the distribution well for both fundamental wave modes. %
This result shows that the method works well for obtaining dispersion characteristics of Lamb waves. %

Although the coupon used here has unidirectional fibre, an orthotropic model was still used for the data at a propagation angle of 90\si{\degree} to test its applicability to all directions. %
\Cref{fig:GFRP_curve_distr_90_deg} indicates that for determining dispersion curves of the $A_0$ mode, this model still provides a useful solution. %
However, the curves generated for the $S_0$ mode become mismatched from the image data, as well as the points taken from the ridge-picking algorithm, at a frequency-thickness greater than 1.2\si{\mega Hz-mm}. %
This indicates that the model used is insufficient for modelling the $S_0$ mode, however, it still generates a reasonable model for the dispersion curves for the $A_0$ mode. %

As stated at the beginning of this paper, a key advantage of the method shown here is the freedom in the posterior distribution, as no assumption is made as to its shape. %
In an engineering context, this allows freedom in the material type to be modelled, so long as the model of the dispersion curve solutions is accurate. %
For both propagation angles here, the univariate distributions of the parameters do not all appear to be of the same shape. %
In fact, the elastic constants and density appear to converge to a Gamma distribution of varying skewness, and the noise parameter appears to converge to a normal distribution. %
In \Cref{fig:GFRP_correl_0_deg,fig:GFRP_correl_90_deg}, all elastic constants appear to converge to Gamma distributions; this indicates that the true posterior of the elastic constants should converge to a skewed distribution. %

\subsection{Quantifying the posterior distributions}

For each of the parameters, the expected value and variance are calculated as the first two arithmetic statistical moments. %
Using a kernel density estimate, the mode of the distributions is also calculated and designated as the \emph{most-likely-estimate}. %
The values were calculated for the samples from the posterior distributions for both angles, the results of which are shown in \Cref{tab:post_values}. %
A notable observation here is the relatively-large discrepancy between the mean values for density; for a propagation angle of 90\si{\degree} it is predicted to be much lower. %
This discrepancy may be as a result of the model having to `counteract' any conflict between the model form being fit and the data values. %

Relative to the mean value, the standard deviations of each parameter are similar, which aligns well with the observed posterior distributions that are seen in \Cref{fig:GFRP_correl_0_deg,fig:GFRP_correl_90_deg}. %
This observation can be interpreted as the level of uncertainty being similar for each parameter, meaning that discrepancies are not confined to a single parameter, but are instead in the \emph{combination} of parameters. %
It is important to note that, in the prior, each parameter is treated as independent, whereas the posterior shows that there is strong co-dependence between the parameter's dispersion-curve solutions. %

\begin{table}[h!]
    \centering
    \begin{tabular}{l | l l l | l l l }
         Parameter & \multicolumn{3}{c|}{0\si{\degree}} & \multicolumn{3}{c}{90\si{\degree}} \\
         (\si{GPa}) & $\mathbb{E}[\theta]$ & $\textrm{MLE}[\theta]$ & $\mathbb{V}[\theta]$ & $\mathbb{E}[\theta]$ & $\textrm{MLE}[\theta]$ & $\mathbb{V}[\theta]$ \\
         \hline
        $C_{11}$   & 34.69   & 26.49 & 265.9    & 26.19   & 20.29 & 150.6    \\
        $C_{13}$   & 6.917   & 3.912 & 21.67    & 20.12   & 14.82 & 103.3    \\
        $C_{33}$   & 15.24   & 11.62 & 51.39    & 25.87   & 19.94 & 149.1    \\
        $C_{55}$   & 3.649   & 2.771 & 3.052    & 2.141   & 1.458 & 1.426   \\
        $\rho$     & 1,320.9 & 1003.0& 396,661  & 874.45  & 594.35& 232,668  \\
        $\sigma$   & 35,578  & 35,149& 6.382e6  & 14,427  & 14,166& 1.576e6
    \end{tabular}
    \caption{Expected value (arithmetic mean), most likely estimate (mode), and variance calculated from the samples from the posterior for each parameter and propagation angle.}
    \label{tab:post_values}
\end{table}

When comparing the results at 0\si{\degree} to those shown in \Cref{tab:gfrp_plate_props}, some further discussion can be made on the result and the advantages of the methodology. %
The results compare reasonably well with the provided values, however, there is still some discrepancy. %
This may be a result of there not being enough observations provided from the higher-frequency range, which was shown in \Cref{sec:disp_curve_sens} to be more sensitive to changes in the elastic constants. %
It is, however, important to note the much less accurate values obtained if one were to choose the most-likely-estimate, in comparison to choosing the mean value. %
This, once more, shows the advantage of estimating the posterior distribution, as opposed to simply determining the most likely elastics constants obtained by maximising \Cref{eq:disp_likeli}. %

% \subsection{Discussion}
% \label{sec:discussion}

\begin{figure}[H]
	\begin{adjustwidth}{-\extralength}{0cm}
    \centering
    \begin{subfigure}{9cm}
        \centering
        \includegraphics[width=8.5cm]{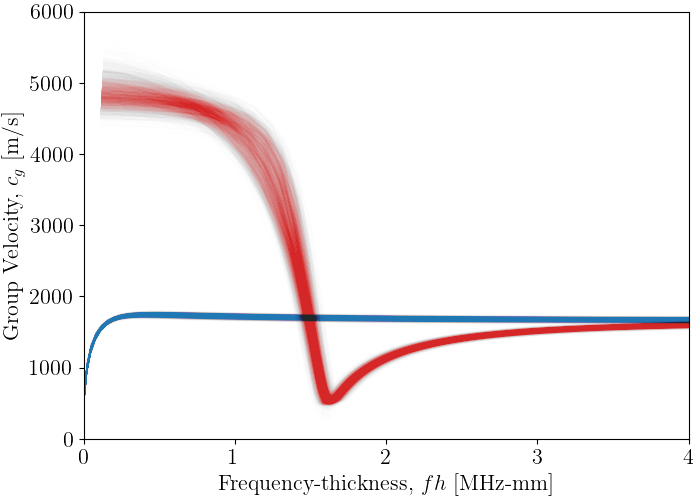}
        \caption{}
        \label{fig:GFRP_cg_distr_0_deg}
    \end{subfigure}
    \begin{subfigure}{9cm}
        \centering
        \includegraphics[width=8.5cm]{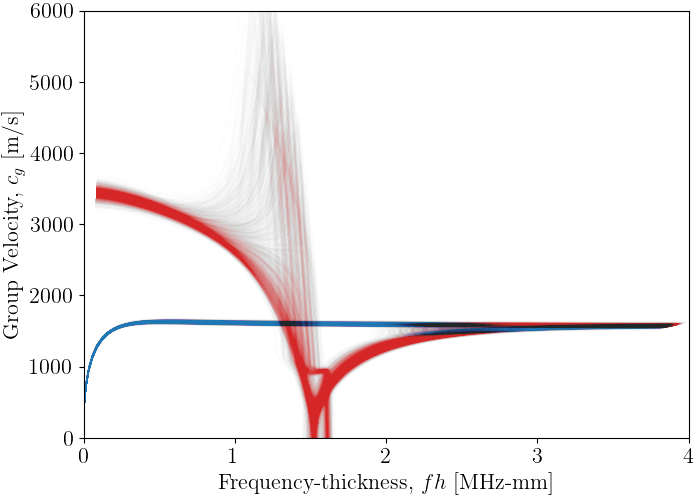}
        \caption{}
        \label{fig:GFRP_cg_distr_90_deg}
    \end{subfigure}
	\end{adjustwidth}
    \caption{Distribution of calculated [$\omega\var{-c_g}$] curves for each sample taken for propagation angles of (a) 0\si{\degree} and (b) 90\si{\degree}. The blue curves show the distribution of the $A_0$ mode and red curves show the distribution of the $S_0$ mode.}
    \label{fig:cg_distrs}
\end{figure}

As the aim of the work here is to determine the dispersion characteristics useful for NDE/SHM strategies, a key motivation of which is to find the group velocity of the waves, it is also useful to look at the distribution of curves for this attribute. %
During the same curve sample-drawing procedure as above, the value of $c_g$ was also calculated as the slope of the generated [$\omega\var{-k}$] curves. %
The distributions of the [$\omega\var{-c_g}$] curves for propagation angles of 0\si{\degree} and 90\si{\degree} are shown in \Cref{fig:cg_distrs}. %
Much like the curves seen in \Cref{fig:GFRP_curve_distrs}, the distribution of the $A_0$ mode is much tighter than that of the $S_0$ mode. %
For the propagation angle of 0\si{\degree}, the curves have no discontinuities and appear to have a low uncertainty. %
% However, the curves for the 90\si{\degree} propagation angle have some discontinuities in the range $1.25<fh<1.8$ MHz-mm. %

In \Cref{fig:GFRP_cg_distr_90_deg}, there are some discontinuities of the curve in the range $1.25<fh<1.8$ MHz-mm. %
From \Cref{eq:wave_relations}, the group velocity is taken as the gradient of the [$\omega\var{-k}$] curves. %
By inspecting \Cref{fig:GFRP_curve_distr_90_deg}, one can see the gradient of the curve changes rapidly, which is the model solution with the highest likelihood. %

From the results shown, the method presented here returns accurate and precise models of the dispersion curves for an arbitrary orthotropic plate. %
The objective problem of the work here was to determine dispersion-curve information on the fundamental modes, as this is the necessary information required for guided wave-based localisation. %
For the purposes of determining the dispersion characteristics from the data provided, it performs well. %

\subsection{On the confidence of the results}

In this paper, the capability of the method has been shown with respect to the initial objective; to determine accurate dispersion curve information for an unknown material, or if the dispersion curve information needs updating. %
It is important to note, that even with confident estimates of these dispersion curves, the estimated variance in each parameter is still reasonably large. %
This indicates that using alignment of estimated dispersion curves, using the most likely estimate of the parameters, is not enough to reasonably state confidence in these parameters. %
In \Cref{sec:disp_curve_sens}, it can be observed that relatively large changes in the material properties result in small changes in the dispersion curve. %
The work here shows the importance of obtaining the posterior distribution of the parameters, as opposed to just the most likely estimate. %

Furthermore, the results for the dispersion curve observations for the 90\si{\degree} data show that the model used here in unsuitable for this propagation direction. %
The model used in the solution equation is for propagation in the direction of the fibres, and these results show there is no combination of material properties that can accurately model dispersion curves for propagation through the fibres. %
Therefore, even though the posterior distribution appear to have settled to a reasonable shape and values, by observing the dispersion curve solutions, one can see the inapplicability of the model. %
These two remarks show the importance of using both the posterior distributions, estimated values/variances, and the dispersion curve solutions to properly assess the results of this procedure. %

%%%%%%%%%%%%%%%%%%%%%%%%%%%%%%%%%%%%%%%%%%
\section{Conclusion}

The aim of the paper was to develop a Bayesian approach to material identification for the purposes of determining dispersion curve models for an orthotropic plate. %
The Bayesian approach allows for inference on the posterior distribution of the material properties, and for total freedom of the distribution shape, as well as inferring any multivariate correlations between parameters. %
By determining the multi-variate posterior distributions of the elastic constant space, it was shown that it is also possible to generate distributions of the dispersion-curve solutions. %
This is another key advantage of the method, as uncertainty bounds on the curve can be propagated through directly to uncertainty in measurements done using these curves -- such as wave source localisation. %
The initial objective for the problem was to determine the dispersion characteristics of the fundamental modes that are important for the purposes of damage localisation in NDE/SHM strategies. %
The results of the curve distributions indicate that the method works well to achieve this objective. %
Future work intended by the authors has been discussed following analysis of the results. %

\subsection{Future work}

As discussed, the method presented here returned reliable and robust results for the objective problem. %
For further application of this method to more exhaustive material identification procedures -- such as full elastic-constant identification or $SH$ dispersion-curve information -- some additions are necessary. %
The key aim for these improvements is to increase the fidelity of the information provided to the procedure in order to allow inference from the additional wave modes. %
An important consideration of this, however, will be the increased computational cost of calculating solutions for more modes. %
The authors intend to explore a number of approaches to address this objective; one such approach is to include rotation of the stiffness matrix combined with using multiple angles for a single observation set, and run a single parameter identification routine. %
Adding further, the Legendre polynomial expansion approach can readily include damping characteristics by using a complex-valued stiffness matrix; by obtaining complex observations of the wavenumber, it would be possible to extend the identification procedure to simulate the posterior distributions on the real-imaginary pairs for each elastic constant. %
Another approach is to develop a multi-dimensional prior definition for the elastic-constant space, which would improve sampling efficiency. %
The final approach to be explored is to adapt the likelihood to use full 2DFT image data, rather than individual observations taken from the image data using a ridge-picking algorithm. %

%%%%%%%%%%%%%%%%%%%%%%%%%%%%%%%%%%%%%%%%%%
\vspace{6pt} 

%%%%%%%%%%%%%%%%%%%%%%%%%%%%%%%%%%%%%%%%%%
%% optional
%\supplementary{The following supporting information can be downloaded at:  \linksupplementary{s1}, Figure S1: title; Table S1: title; Video S1: title.}


%%%%%%%%%%%%%%%%%%%%%%%%%%%%%%%%%%%%%%%%%%
\authorcontributions{
Conceptualization, Marcus Haywood-Alexander and Timothy Rogers; %
Data curation, Marcus Haywood-Alexander and Robin Mills; %
Formal analysis, Marcus Haywood-Alexander; %
Funding acquisition, Nikolaos Dervilis, Keith Worden and Purim Ladpli; %
Investigation, Marcus Haywood-Alexander; %
Methodology, Marcus Haywood-Alexander, Nikolaos Dervilis and Timothy Rogers; %
Project administration, Nikolaos Dervilis and Keith Worden; %
Resources, Robin Mills and Purim Ladpli; %
Software, Marcus Haywood-Alexander and Timothy Rogers; %
Supervision, Nikolaos Dervilis, Keith Worden and Timothy Rogers; %
Validation, Marcus Haywood-Alexander and Timothy Rogers; %
Visualization, Marcus Haywood-Alexander; %
Writing - original draft, Marcus Haywood-Alexander and Timothy Rogers; %
Writing - review \& editing, Nikolaos Dervilis, Keith Worden, Robin Mills and Purim Ladpli. %
All authors have read and agreed to the published version of the manuscript.}

\funding{This research was funded by the UK Engineering and Physical Sciences Research Council (EPSRC), grant numbers grant numbers EP/R004900/1, EP/R003645/1 and EP/N010884/1.}

\abbreviations{Abbreviations}{
The following abbreviations are used in this manuscript:\\

\noindent 
\begin{tabular}{@{}ll}
MCMC & Markov-Chain Monte Carlo\\
HMC & Hamiltonian Monte Carlo\\
NDE & Non-Destructive Evaluation\\
SHM & Structural Health Monitoring\\
UGW & Ultrasonic Guided Wave\\
GFRP & Glass-Fibre Reinforced Polymer\\
LPE & Legendre Polynomial Expansion\\
2DFT & Two-Dimensional Fourier Transform\\
PZT & Piezo-electric Transducer
\end{tabular}
}

%%%%%%%%%%%%%%%%%%%%%%%%%%%%%%%%%%%%%%%%%%
%% Optional
\appendixtitles{no} % Leave argument "no" if all appendix headings stay EMPTY (then no dot is printed after "Appendix A"). If the appendix sections contain a heading then change the argument to "yes".
\appendixstart
\appendix
\section[\appendixname~\thesection]{}
\label{app:orth_disp_mat}
\Cref{sec:orth_disp_solve} shows the method of solving dispersion curves for anisotropic materials using the Legendre polynomial expansion approach. %
The final system of equations which forms an eigenvalue problem are given in \Cref{eq:orth_disp_final_eq}, the matrix elements of which are given below,

\begin{equation}
    M_{jm} = NT_1(m,j,0)
\end{equation}
\begin{equation}
    A_{11}^{jm} = -\left(\frac{C_{11}}{\rho}\right)NT_1(m,j,0) + \left(\frac{C_{55}}{\rho}\right)NT_1(m,j,2) + \left(\frac{C_{55}}{\rho}\right)NT_2(m,j,1)
\end{equation}
\begin{equation}
    A_{13}^{jm} = i\left(\frac{C_{13}+C_{55}}{\rho}\right)NT_1(m,j,1) + i\left(\frac{C_{55}}{\rho}\right)NT_2(m,j,0)
\end{equation}
\begin{equation}
    A_{31}^{jm} = i\left(\frac{C_{31}+C_{55}}{\rho}\right)NT_1(m,j,1) + i\left(\frac{C_{31}}{\rho}\right)NT_2(m,j,0)
\end{equation}
\begin{equation}
    A_{33}^{jm} = -\left(\frac{C_{55}}{\rho}\right)NT_1(m,j,0) + \left(\frac{C_{33}}{\rho}\right)NT_1(m,j,2) + \left(\frac{C_{33}}{\rho}\right)NT_2(m,j,1)
\end{equation}
\begin{equation}
    NT_1(m,j,n) = \int^{kh}_0 Q_j^*(q_3)\frac{\partial^n}{\partial q_3^n}Q_m(q_3)\;dq_3
\end{equation}
\begin{equation}
    NT_2(m,j,n) = \int^{kh}_0 Q_j^*(q_3)[\delta(q_3=0)-\delta(q_3=kh)]\frac{\partial^n}{\partial q_3^n}Q_m(q_3)\;dq_3
\end{equation}
\begin{equation}
    Q_m(q_3) = \sqrt{\frac{2m+1}{kh}}P_m(\tilde{q}_3),\qquad \tilde{q}_3 = \frac{2q_3}{kh} - 1
\end{equation}

\section[\appendixname~\thesection]{}
\label{app:num_tricks}
A number of numerical manipulations were employed here to reduce computational cost, the details of which will be outlined here.

\subsection[\appendixname~\thesubsection]{}
Symbolic mathematical programming is very computationally expensive, but the Legendre expansion approach produces only polynomial equations. %
These polynomial equations can be generated and manipulated symbolically. %
A particular manipulation can be found in implementing the calculation of $NT_2(m,j,n)$,
\begin{equation}
    NT_2(m,j,n) = \int^{kh}_0 Q_j^*(q_3)[\delta(q_3=0)-\delta(q_3-kh)]\frac{\partial^n}{\partial q_3^n}Q_m(q_3)\;dq_3
\end{equation}
\begin{equation}
    NT_2(m,j,n) = \int^{kh}_0 f_n(q_3)\delta(q_3)dq_3  - \int^{kh}_0 f_n(q_3)\delta(q_3-kh)\;dq_3
\end{equation}
where,
\begin{equation}
    f_n(q_3) = Q_j^*(q_3)\frac{\partial^n}{\partial q_3^n}Q_m(q_3)
\end{equation}
and the sifting property of a Dirac delta function states,
\begin{equation}
    \int^\infty_{-\infty} f(x)\delta(x-x_0) dx = f(x_0)
\end{equation}
which leads to,
\begin{equation}
    NT_2(m,j,n) = f_n(0) - f_n(kh)
\end{equation}

\subsection[\appendixname~\thesubsection]{}
During solution of the dispersion-curve equations, the matrix A is produced. %
This matrix is formed from 4 sub-matrices $A_{11}$, $A_{31}$, $A_{13}$ and $A_{33}$, all of size $2(M+1) \times 2(M+1)$, 
\begin{equation}
    A = \begin{bmatrix}
        A_{11} & A_{13} \\
        A_{31} & A_{33}
    \end{bmatrix}
\end{equation}
where $A_{11},\; A_{33} \in \mathbb{R}^{\hat{M},\hat{M}}$ and $A_{13},\; A_{31} \in \mathbb{C}^{\hat{M},\hat{M}}$ where $\textrm{Re}(A_{13}) = \textrm{Re}(A_{31}) = 0$. %
The use of complex numbers greatly increases computational cost. %
However, by some manipulation, the requirement for complex variable types could be eradicated, decreasing computational load. %
The eigendecomposition problem is formulated such that,
\begin{equation}
    |A - \lambda\mathbf{I}| = 0
\end{equation}
where $\hat{M}=M+1$, $\lambda_1 = \lambda\{1:\hat{M}\}$ and $\lambda_2 = \lambda\{\hat{M}+1:2\hat{M}\}$,
\begin{equation}
    \begin{vmatrix}
        A_{11} - \lambda_1\mathbf{I}_{\hat{M}} & A_{13} \\
        A_{31} & A_{33} - \lambda_2\mathbf{I}_{\hat{M}}
    \end{vmatrix} = 0
\end{equation}
\begin{equation}
    \begin{vmatrix}
        A_{11} - \lambda_1\mathbf{I}_{\hat{M}} & A_{13} \\
        A_{31} & A_{33} - \lambda_2\mathbf{I}_{\hat{M}}
    \end{vmatrix}
     = |A_{11} - \lambda_1\mathbf{I}_{\hat{M}}||A_{33} - \lambda_2\mathbf{I}_{\hat{M}}| - |A_{13}||A_{31}|
\end{equation}
as $A_{13},A_{31}$ are purely imaginary, the following manipulation is then applied,
\begin{equation}
    |A_{13}||A_{31}| = |i\textrm{Im}(A_{13})||i\textrm{Im}(A_{31})| = |-\textrm{Im}(A_{13})||\textrm{Im}(A_{31})|
\end{equation}
and thus the matrix $A$ can be reformulated as,
\begin{equation}
    \hat{A} = \begin{bmatrix}
        A_{11} & -\textrm{Im}(A_{13}) \\
        \textrm{Im}(A_{31}) & A_{33}
    \end{bmatrix}
\end{equation}
\begin{equation}
    |\hat{A} - \lambda\mathbf{I}| = 0
\end{equation}

\subsection[\appendixname~\thesubsection]{}
The returned eigenvectors $\lambda$ of the matrix $A$, give solutions for $2\hat{M}$ modes where,
\begin{equation}
    \lambda_i = \{\lambda_1,\lambda_2,...,\lambda_{2\hat{M}}\}
\end{equation}
Each eigenvalue can be related to the phase velocity of each mode by,
\begin{equation}
    \lambda_i = -\frac{\omega_j^2}{k_j^2} = -c_{p_j}^2
\end{equation}
where $j = 2\hat{M} - i + 1$. %
The only physically-possible solutions are $c_p \in \mathbb{R}_{>0}$; therefore, only solutions where $\lambda \in \mathbb{R}_{<0}$ are viable. %
The minimum-magnitude eigenvalue will represent the smallest-value solution of $c_p$. %
As here, data points for the $A_0$ mode are being used, only the smallest value of $c_p$ is needed (and thus smallest-magnitude eigenvalue). %
This leads to the use of the power iteration method \cite{Mises1929}, which is a method that returns only the dominant eigenvalue of a matrix, at reduced computational expense. %
Unfortunately, the dominant eigenvalue of a matrix $A$ is the largest magnitude, which would return the largest value of $c_p$. %
However, the smallest magnitude eigenvalue can be returned by determining the dominant eigenvalue of $A^{-1}$. %

%%%%%%%%%%%%%%%%%%%%%%%%%%%%%%%%%%%%%%%%%%
\begin{adjustwidth}{-\extralength}{0cm}

\reftitle{References}


\bibliography{biblio}

\end{adjustwidth}
\end{document}